
\magnification=1200
\hsize=13.5cm

\def\sk{\vskip .4cm}
\def\noi{\noindent}
\def\om{\omega}

\def\del{\delta}

\def\unmezzo{{1 \over 2}}
\def\epsi{\varepsilon}
\def\we{\wedge}

\def\part{\partial}

\def\pdxi{{\partial \over {\partial x^i}}}

\def\qP{q-Poincar\'e~}
\def\R#1#2{ {\hat R}^{#1}_{~~~#2} }
\def\Rt#1{ {\hat R}_{#1} }
\def\Rh{\hat R}
\def\C#1#2{ {\bf C}^{#1}_{~#2} }
\def\c#1#2{ C^{#1}_{~#2} }
\def\q#1{   {{q^{#1} - q^{-#1}} \over {q^{\unmezzo}-q^{-\unmezzo}}}  }
\def\DR{\Delta_R}
\def\DL{\Delta_L}
\def\Rh{{\hat R}}
\def\f#1#2{ f^{#1}_{~~#2} }
\def\F#1#2{ F^{#1}_{~~#2} }
\def\M#1#2{ M_{#1}^{~#2} }
\def\qm{q^{-1}}
\def\um{u^{-1}}
\def\vm{v^{-1}}
\def\xm{x^{-}}
\def\xp{x^{+}}
\def\fm{f_-}
\def\fp{f_+}
\def\fn{f_0}
\def\D{\Delta}
\def\Mat#1#2#3#4#5#6#7#8#9{\left( \matrix{
     #1 & #2 & #3 \cr
     #4 & #5 & #6 \cr
     #7 & #8 & #9 \cr
   }\right) }
\def\Ap{A^{\prime}}
\def\Dp{\Delta^{\prime}}
\def\ep{\epsi^{\prime}}
\def\kp{\kappa^{\prime}}
\def\Cb{{\bf C}}

\hskip 10cm \vbox{\hbox{DFTT-56/91}\hbox
{December 1991}}
\vskip 0.6cm
\centerline{\bf BICOVARIANT DIFFERENTIAL CALCULUS}
\centerline{\bf ON THE QUANTUM D=2 POINCAR\'E GROUP}
\vskip 3cm
\centerline{\bf Leonardo Castellani}
\vskip .4cm
\centerline{\sl Istituto Nazionale di Fisica Nucleare, Sezione di Torino}
\centerline{\sl Via P. Giuria 1, I-10125 Torino, Italy}
\vskip 3cm
\centerline{\bf Abstract}
\vskip .5cm

We present a bicovariant differential calculus on the quantum Poincar\'e
group in two dimensions. Gravity theories on quantum groups are
discussed.

\vskip 5cm
\vbox{\hbox{DFTT-56/91}\hbox{December 1991}}

\vfill
\eject

Quantum groups, or q-deformed function algebras of classical groups [1],
have recently appeared to be the underlying algebraic structures of some
two-dimensional and three-dimensional quantum field theories. In this
paper we will be concerned with the geometry of quantum groups, and
more precisely with the differential geometry of the D=2 \qP group.
\sk

Our objective is to construct a q-deformed gravity as a ``gauge" theory
of the D=4 \qP group, which reduces for q=1 to ordinary gravity. The D=2
case serves both as a preliminary step and as a differential geometric
setting for D=2 q-gravity.
\sk

Why should we want to deform ordinary gravity (or supergravity,
superstrings etc.)? Continuous deformations of ``classical" theories
often have physical relevance: consider e.g. the deformation
parameters $v/c$, $\hbar$ or the De Sitter radius. Also, for distances
of the order of the Planck length the smoothness of spacetime is really
a mathematical assumption, leading to infinities in quantum gravity. A
generalization of Riemannian geometry to ``something else" is perhaps
necessary, and the geometry of quantum groups seems to
offer a way, leading to non-commutative spaces with group structure.
For a review on non-commutative geometry (not necessarily linked to
quantum groups) and its potential uses in physics, see e.g. ref. [2].
\sk

The geometrical aspects of quantum groups have been very recently
investigated by a number of researchers [3,4,5,6,7,8], from
two main viewpoints:

i) as the (noncommutative) geometry of a representation space
for the quantum group action [4,5,6];

ii) as the (noncommutative) geometry of the group space itself [3,7,8].

The construction of a
bicovariant \footnote{*}{i.e. covariant under the left and right actions
of the group, see later.} differential calculus
on quantum groups has been initiated
by Woronowicz in [3], and applied to q-SU(2) [3a] and to the q-Lorentz
group [8]. A general framework for the general cases of q-SO(N) and
q-SU(N) has been proposed in [7].
\sk
Here we use most of the tools developed in ref.s [3], and present a
bicovariant differential calculus for the two-dimensional \qP group.
An analogous result for the D=4 \qP group will be presented elsewhere
 [9].
\sk

Following the general formulation of [3c], we consider the q-commutator
algebra:

$$T_i T_j - \R{kl}{ij} T_k T_l = \C{k}{ij} T_k \eqno(1) $$

\noi where:
\sk

i) $T_i$ are linear functionals on the algebra $A$ of ``functions on the
quantum group", ~~~~the q-analogue of the Lie algebra generators for
ordinary groups; they belong to the algebra $\Ap$, the dual of $A$. $A$
and $\Ap$ are Hopf algebras with dual Hopf structures: the product in
$\Ap$, appearing in (1), is defined by

$$T_i T_j (a) \equiv (T_i \otimes T_j) \Delta (a) \eqno(2) $$

\noi with $\Delta = $ coproduct in $A$. The
unit of $\Ap$ is the counit of $A
$ and so on. To fix our notations:

- the coproduct, counit and coinverse of $A$ are denoted respectively
  $\Delta$, $\epsi$ and $\kappa$. The unit of $A$ is $I$.

- the coproduct, counit and coinverse of $\Ap$ are denoted respectively
  $\Dp$, $\ep$ and $\kp$. The unit of $\Ap$ is $1$.
\sk

ii) $\R{kl}{ij}$ is the braiding matrix satisfying

$$\R{kl}{ij} \R{lm}{sp} \R{ks}{qu}=\R{jm}{kl} \R{ik}{qs} \R{sl}{up}
\eqno(3)$$

or, in more compact notation

$$ \Rt{12} \Rt{23} \Rt{12} = \Rt{23} \Rt{12} \Rt{23}  $$

iii) $\C{k}{ij}$ are the q-structure constants satisfying the q-analogue
of Jacobi identities:

$$\C{n}{il} \C{l}{jk} = \C{l}{ij} \C{n}{lk} - \R{lm}{jk} \C{r}{il}
\C{n}{rm} \eqno(4) $$
\sk

A bicovariant differential calculus on the quantum group associated to
(1) exists when, besides (3) and (4), two additional conditions are
satisfied (see later):

$$\C{j}{is} \R{sq}{rl} \R{ir}{pk} + \C{q}{rl} \R{jr}{pk} = \R{jq}{ri}
\R{si}{kl} \C{r}{ps} + \R{jq}{pi} \C{i}{kl} \eqno(5)$$

$$\C{j}{ef} \R{ie}{bq} \R{qf}{ca} = \R{ij}{da} \C{d}{bc} \eqno(6)$$

The four relations (2), (3), (4) and (5) are henceforth called the
{\sl bicovariance conditions}.
Note that the classical limit of $\R{ij}{kl}$ is $\delta^i_l
\delta^j_k$ so that relations (3), (5) and (6) become trivial in this
limit.
\sk

An example where (3) is satisfied, but (4), (5) and (6) are not, is
given
by the q-SU(2) algebra:

$$ q^{-1} T_0 T_+ - q T_+ T_0 = T_+ $$
$$ q T_0 T_- - q^{-1} T_- T_0 = -T_- $$
$$ q^{\unmezzo} T_+ T_- - q^{-\unmezzo} T_- T_+ = T_0  \eqno(7) $$

\noi obtained from the Drinfeld-Jimbo q-SU(2) [1] :

$$[H_0, H_{\pm} ]= \pm H_{\pm}, ~~[H_+, H_-] = \unmezzo \q{H_0}
\eqno(8)$$

\noi via the mapping [10]

$$T_0 = { {1 - q^{-2 H_0}} \over {q-q^{-1}}},~~T_{\pm} = q^{-{H_0 \over
2}} \sqrt{{2 \over {q^{\unmezzo} + q^{-\unmezzo}}}}~H_{\pm} \eqno(9) $$

Indeed the differential calculus corresponding to (7) is only {\sl left
covariant} (and is essentially equivalent to the left covariant
differential calculus on the q-SU(2) of ref. [3a]).
\sk
Let us recall the basic facts about bicovariant differential calculus on
quantum groups. It is defined by [3c] :
\sk

1) a linear map $d~:~A \rightarrow \Gamma$, satisfying the Leibniz rule
$d(ab)=(da)b+a(db),~\forall a,b~ \in A$. $\Gamma$ is a bimodule over
$a$, which essentially means that it can be multiplied from the left and
from the right by elements of $A$, and is the q-analogue of the space
of 1-forms on a Lie group. Every element $\rho$ of $\Gamma$ is assumed
to be expressible as $\rho = \sum_k a_k db_k $.
\sk

2) the {\sl left action} of the quantum group on $\Gamma$:

$$\Delta_L ~:~\Gamma \rightarrow A \otimes \Gamma,~~~~\Delta_L (\sum_k
a_k db_k)=\sum_k \Delta (a_k) (id \otimes d)\Delta (b_k)  \eqno(10) $$

\noi generalizes the ordinary pull-back on forms. If we call $L_x$ the
left multiplication by an element $X$ of a group $G$:

$$L_x y \equiv xy~~~\forall x,y \in G $$

\noi then the pull-back $L_x^*$ on 1-forms $\om$ is defined by:

$$(L_x^* \om)(y)=\om(xy)|_y $$

\noi and we can consider the map $L^*$:

$$ L^* : T^* G  \rightarrow Fun(G) \times T^* G$$
$$L^* \om (x,y) \equiv (L^*_x \om)(y) $$

The left action $\Delta_L$ reduces to $L^*$ in the classical limit. The
 map $\Delta_L$ is assumed to satisfy properties that q-generalize
the well-known properties of $L^*$ (see [11] for a detailed discussion).

Left-invariant 1-forms $\om$ classically satisfy $L^*_x \om=\om$ or
equivalently:

$$L^* \om=I \otimes \om        \eqno(11)$$

\noi where $I$ is the identity in $A$, i.e. the function on $G$
that sends all elements of $G$ into the identity of ${\bf C}$. Indeed

$$L^* \om (x,y) = (I \otimes \om ) (x,y)= I(x) \om(y) = \om (y)$$

In complete analogy, left-invariant 1-forms $\om$ on quantum groups
are defined to satisfy

$$\Delta_L (\om) = I \otimes \om  \eqno(12)$$

\sk

3) the {\sl right action} of the quantum group on $\Gamma$:

$$\Delta_R : \Gamma \rightarrow \Gamma \otimes A,~~~\Delta_R
(\sum_k a_k db_k)=\sum_k
\Delta (a_k) (d \otimes id) \Delta (b_k)  \eqno(13)$$

\noi generalizes the ordinary pull-back on forms induced by right
multiplication $R_x y = y x$. The discussion in 2) can be repeated for
$\Delta_R$.
\sk

4) the q-analogue of the fact that left and right actions commute
($L^*_x R^*_y = R^*_y L^*_x$) :

$$ (id \otimes \Delta_R) \Delta_L = (\Delta_L \otimes id) \Delta_R
    \eqno(14)$$

We now summarize some important consequences of 1), 2), 3) and  4),
derived in [3c].
\sk

As in the classical case, the whole of $\Gamma$ is generated by $\{ \om^
i \}$, a basis in the vector space $\Gamma_{inv}$ of all left-invariant
elements of $\Gamma$ (or by $\{ \eta^i\}$, a basis for the right-
invariant elements of $\Gamma$, for which $\DR (\eta^i) = \eta^i \otimes
I$) Any $\rho \in \Gamma$ is expressible (uniquely) as $\rho = a_i
\om^i$ or also as $\rho = \om^i b_i$. Therefore we must have:

$$\om^i b = F^i_{~j} (b) \om^j, ~~~~F^i_{~j} :A \rightarrow A ~~{\rm
   linear~map} $$

\noi or equivalently

$$\om^i b= (\f{i}{j} \star b) \om^j \equiv (id \otimes \f{i}{j})
\Delta (b)
\om^j \eqno(15)$$

\noi where $\f{i}{j}$ $\in \Ap$ is the functional
on $A$ defined by $\f{i}{j} (a)
= \epsi (\F{i}{j} (a))$, $\epsi$ being the counit of $A$. The same
reasoning for right-invariant 1-forms $\eta^i$ yields the equation:

$$b\eta^i=\eta^j (b \star (\f{i}{j} \circ \kappa)) \equiv \eta^j (
(\f{i}{j} \circ \kappa) \otimes id) \Delta (b),~~~\kappa = {\rm
coinverse~in~}A $$

For consistency, the functionals $\f{i}{j}$ must satisfy the relations:

$$\f{i}{j} (ab)= \sum_k \f{i}{k} (a) \f{k}{j} (b) \eqno(16a)$$
$$\f{i}{j} (I) = \del^i_j \eqno(16b)$$
$$(\f{k}{j} \circ \kappa ) \f{j}{i} = \del^k_i ~1;~~~
   \f{k}{j} (\f{j}{i} \circ \kappa) \f{j}{k} = \del^k_i ~1;~~~
   \eqno(16c)$$

\noi so that their coproduct, counit and coinverse are given by:

$$\Dp (\f{i}{j})=\f{i}{k} \otimes \f{k}{j}   \eqno(17a)$$

$$\ep (\f{i}{j}) = \del^i_j  \eqno(17b)$$

$$\kp (\f{i}{j})= \f{i}{j} \circ \kappa  \eqno(17c)$$
\sk

The {\sl quantum group element in the adjoint representation} can be
introduced as follows. It is easy to show that $\DR (\om^i)$
belongs to $\Gamma_{inv} \otimes A$, and therefore:

$$ \DR (\om^i) = \om^j \otimes \M{j}{i},~~~\M{j}{i} \in A; \eqno(18)$$

\noi in the classical case, $\M{j}{i}$ is the adjoint representation of
$G$. This justifies our calling $\M{j}{i}$ the adjoint representation of
the quantum group.
The coproduct, counit and coinverse of $\M{j}{i}$ can be deduced to be
[3c]:

$$\Delta (\M{j}{i}) = \M{j}{l} \otimes \M{l}{i}, ~~~\epsi (\M{j}{i}) =
\delta^i_j $$

$$\kappa (\M{i}{l}) \M{l}{j}=\delta^j_i=\M{i}{l} \kappa (\M{l}{j})
   \eqno(19)$$

\noi Moreover one can prove the relation:

$$\M{i}{j} (a \star \f{i}{k})=(\f{j}{i} \star a) \M{k}{i} \eqno(20)$$

\sk
The last ingredient we need is the {\sl exterior product}. It is defined
by an automorphism $\Rh$ in $\Gamma \otimes \Gamma$ that generalizes the
ordinary permutation operator:

$$\Rh (\om^i \otimes \eta^j )= \eta^j \otimes \om^i  $$

i) in general $\Rh^2 \not= 1$, since $\Rh (\eta^j \otimes \om^i )$ is not
necessarily equal to $ \om^i \otimes \eta^j $. By linearity $\Rh$ can
be extended to the whole of $\Gamma \otimes \Gamma$.

ii) $\Rh$ is invertible and commutes with the left and right actions of
$G$. $\Rh (\om^i \otimes \om^j)$ is left-invariant, so that

$$\Rh (\om^i \otimes \om^j)= \R{ij}{kl} \om^k \otimes \om^l  \eqno(21) $$

iii) we have

$$ \R{ij}{kl} = \f{i}{l} (\M{k}{j});           \eqno(22) $$

\noi thus the quantities $\f{i}{l}$, $\M{k}{j}$ characterizing the
bimodule $\Gamma$ are dual in the sense of eq. (22) and determine the
exterior product:

$$\rho \we \rho ' \equiv \rho \otimes \rho ' - \Rh (\rho \otimes \rho ')
$$

$$\om^i \we \om^j \equiv \om^i \otimes \om^j - \R{ij}{kl} \om^k \otimes
\om^l  \eqno(23) $$
\sk

Having the exterior product we can define the {\sl exterior
differential}

$$d~:~\Gamma \rightarrow \Gamma \we \Gamma$$

$$d(\sum_k a_k db_k) = da_k \we db_k~~~( \Rightarrow d^2 =0)$$

\noi which can easily be extended to
$\Gamma^{\we n}$ ( $d: \Gamma^{\we n}
\rightarrow \Gamma^{\we (n+1)}$) and can be shown to commute with
the left and right actions
of $G$, as in the classical case [3c].
\sk

Finally, the hypotheses 1), 2), 3), 4) imply the existence of the $T_i$
functionals satisfying eq. (1), and of the corresponding Cartan-Maurer
equations for the left-invariant forms $\om^i$ [3c]. In eq. (1)
the $\Rh$ matrix is the same that defines the wedge product in (23).
One can prove that the $\Cb$ and $\Rh$ tensors do indeed satisfy the
four conditions (3), (4), (5), (6). The action of $T_i$ on
$\M{j}{k}$ is found to be:

$$ T_i (\M{k}{l}) = \C{l}{ki} \eqno(25)$$

In the sequel we also need the relations [3c,12,11]

$$ T_i(ab)=\Dp (T_i) (a \otimes b )   \eqno(26)$$

$$ da=(T_i * a) \om^{i}    \eqno(27)$$

\noi See for ex. [11] for a detailed discussion on how
to obtain (5) and (6), and
eq. (25); all these relations are already implicitly contained in [3c]
(see also [12]).
\sk
Using (22) and (25), we see that the conditions (3), (5) and (6)
can be cast into operator form as:

$$\R{ab}{nm} \f{n}{c} \f{m}{d} = \f{a}{n} \f{b}{m} \R{nm}{cd} \eqno(28a)
$$

$$ \f{a}{c} T_b + \C{a}{nm} \f{n}{c}  \f{m}{b} = \f{a}{n} \C{n}{cb} +
T_n \f{a}{m} \R{nm}{cb} \eqno(28b)$$

$$ T_k \f{n}{l} = \R{ij}{kl} \f{n}{i} T_j \eqno(28c)$$

\noi whereas condition (4) follows from the definition of
the q-commutations (1) applied to $\M{i}{j}$. In
fact, the bicovariance conditions
define a {\sl quasi triangular quantum Lie algebra}, whose associated
quantum group admits a bicovariant differential calculus \footnote{*}
{Strictly speaking, (3), (5) and (6) do not necessarily imply
that eqs. (28)
hold on the whole algebra $A$, since the algebra generated by the
$\M{i}{j}$ is in general a subalgebra of $A$.}(see e.g.[12]).
By ``associated quantum group" we mean the Hopf algebra with Hopf
structures dual to those of the quantum Lie algebra.
At this juncture we note that the coproduct, counit and coinverse of the
q-algebra (1) can be consistently defined as :

$$ \Dp (T_i) = T_j \otimes \f{j}{i} + 1 \otimes T_i  \eqno(29a)$$

$$ \ep (T_i) = 0 \eqno(29b)$$

$$ \kp (T_i) = - T_j  ~\kp (\f{j}{i})  \eqno(29c) $$

\sk

Viceversa, suppose that we find some tensors $\R{ij}{kl}$ and $\C{i}{jk}$
satisfying the bicovariance conditions (3), (4), (5), (6).
Then we have a quasi triangular quantum Lie
algebra, and we can construct a bicovariant calculus
on the quantum group dual to the q-commutator algebra (1). We illustrate
the method in the example of D=2 \qP. Consider the 3-parameter deformed
algebra:

$$T_0 T_+ - T_+ T_0 = r T_+$$

$$T_0 T_- - T_- T_0 = -s T_-$$

$$q^{\unmezzo} T_+ T_- - q^{-\unmezzo} T_- T_+ = 0   \eqno(30)$$

\noi which reduces to the D=2 Poincar\'e algebra for $r \rightarrow 1$,
$s \rightarrow 1$ and $q \rightarrow 1$.
The corresponding $\Rh$ and ${\bf C}$ tensors are

$$\R{-+}{+-} = \qm = \f{-}{-} (\M{+}{+}),~~~~\R{+-}{-+}=q=\f{+}{+}
  (\M{-}{-})$$
$$\R{+0}{0+} = 1 = \f{+}{+} (\M{0}{0}),~~~~\R{0+}{+0}=1=\f{0}{0}
  (\M{+}{+})  $$
$$\R{-0}{0-} = 1 = \f{-}{-} (\M{0}{0}),~~~~\R{0-}{-0}=1=\f{0}{0}
  (\M{-}{-})$$
$$\R{++}{++} = 1 = \f{+}{+} (\M{+}{+}),~~~~\R{--}{--}=1=\f{-}{-}
  (\M{-}{-}),~~~~\R{00}{00} = 1 = \f{0}{0} (\M{0}{0}) \eqno(31) $$

\sk

$$\C{-}{-0}=s=T_0(\M{-}{-}),~~~~\C{+}{+0}= -r = T_0 (\M{+}{+})$$
$$\C{-}{0-}=-s=T_-(\M{0}{-}),~~~~\C{+}{0+}= r = T_+ (\M{0}{+}) \eqno(32)$$

\noi all other components vanishing.

It is not difficult to check that the bicovariance conditions (3), (4)
(5) and (6) are indeed fulfilled by the tensors given above. In (29)
and (30) we have adjoined the equalities due to eqs. (22) and (25).
The matrix  $\M{i}{j}$ can be seen as the generic quantum group
element in the adjoint representation (cfr. eq. (18)). Let us try to
give this matrix an explicit form. To get a feeling of what it should
look like, we consider it in the limit q=1. In the classical case
the Lie algebra generators in the adjoint representation are

$$ (T_0)^i_{~j} = -\C{i}{0j} = \Mat{0}{0}{0}{0}{-1}{0}{0}{0}{1} $$

$$ (T_+)^i_{~j} = -\C{i}{+j} = \Mat{0}{1}{0}{0}{0}{0}{0}{0}{0} $$

$$ (T_-)^i_{~j} = -\C{i}{-j} = \Mat{0}{0}{-1}{0}{0}{0}{0}{0}{0}
\eqno(33)$$

The generic group element of D=2
Poincar\'e in the adjoint representation can then be
parametrized with exponential coordinates :

$$M = \exp (x^0T_0) \exp (x^+ T_+) \exp (x^- T_-)  =
    \Mat{1}{x^+}{-x^-}{0}{\exp (-x^0)}{0}{0}{0}{\exp (x^0)} \eqno(34)
$$

\noi as the reader can verify by using eqs. (33).
\sk

In the quantum case, the matrix elements $\M{i}{j}$ do not commute any
more, but satisfy the exchange relations:

$$\M{i}{j} \M{r}{q} \R{ir}{pk} = \R{jq}{ri} \M{p}{r} \M{k}{i} \eqno(35)
$$

\noi that are obtained by taking $a=\M{p}{q}$ in eq.(20),
and using (22). To obtain further information on the elements of the
quantum $\M{i}{j}$, we apply the formula (27) to $\M{i}{j}$ :

$$d \M{i}{j}=(T_k \star \M{i}{j} ) \om^k \equiv (id \otimes T_k)
\Delta (\M{i}{j})=\M{i}{l} T_k (\M{l}{j})  \eqno(36)$$

\noi and find the differentials:

$$d\M{0}{0}=d\M{+}{0}=d\M{-}{0}=0$$
$$d\M{+}{+}=r(-\M{+}{+} \om^0+\M{+}{0} \om^+),
{}~~~d\M{-}{-}=s(\M{-}{-} \om^0-\M{-}{0} \om^-) $$
$$d\M{+}{-}=s(\M{+}{-} \om^0-\M{+}{0} \om^-),
{}~~~d\M{-}{+}=r(-\M{-}{+} \om^0+\M{-}{0} \om^+) $$
$$d\M{0}{+}=r(-\M{0}{+} \om^0+ \om^+),
{}~~~d\M{0}{-}=s(-\M{0}{-} \om^0+ \om^-) \eqno(37)$$

\noi From the first line we see that $\M{0}{0}$, $\M{+}{0}$ and $\M{-}{0}$
must be constants (i.e. proportional to the unit of $A$); moreover from
eq. (35) we find that $\M{+}{0} \M{-}{0}=\qm \M{-}{0} \M{+}{0}$, which implies
that the constant $\M{+}{0}$, $\M{-}{0}$ must vanish. The element $\M{0}{0}$,
which according to (35) commutes with all the other elements, can be chosen
equal to the unit $I$. Finally, the elements $\M{+}{-}$ and $\M{-}{+}$ can
also consistently be taken as vanishing, and we arrive at the quantum
matrix:

$$\M{i}{j} = \Mat{I}{{1 \over r}x^+}{-{1 \over s}x^-}{0}
    {u}{0}{0}{0}{v} \eqno(38)$$

\noi whose elements satisfy the commutations:

$$x^+ x^- = q x^- x^+  $$
$$u x^-= q x^- u $$
$$v x^+ = \qm x^+ v \eqno(39)$$

\noi all other being trivial. The ``coordinates" $x^i$ belonging to $A$
are defined by:

$$T_i (x^j)=\del^j_i  \eqno(40)$$

\noi (such $x^i~ \in~A$ can always be found, see [3c]).

\sk
{\sl Note 1}: the commutation relation of the quantum plane coordinates
$x^+$ and $x^-$ is formally
similar to the one of ref.s [4,5]. However here we are dealing with light-cone
coordinates.
\sk

{\sl Note 2}: it is not obvious how one can parametrize the diagonal elements
$u$ and $v$ with the coordinate $x^0$.
\sk

{\sl Note 3}: as in the classical case, $T_i$ can be represented as
${\part \over {\part x^i}} |_{x=0}$, the partial derivatives being defined by
$da=(T_i \star a)\om^i \equiv  (\pdxi a) \om^i ,~~\forall a \in A$.
\sk

{\sl Note 4}: the product of two matrices $M(x,u,v)$ and $M(y,w,z)$ of
type (34) yields a matrix $M$ of the same type:

$$M = \Mat{1}{{1 \over r}(y^+ + x^+ w)}{-{1 \over s}(y^- +x^- z)}
               {0}{uw}{0}
               {0}{0}{vz} \eqno(41)$$

The inverse
(in the matrix sense) of (34) gives the explicit form of the coinverse
of $\M{i}{j}$, since $\kappa (\M{i}{j} ) \M{j}{k} = \del^k_i $ (cfr. eq.
(19)):

$$(M^{-1})_i^{~j} = \kappa (\M{i}{j}) =
\Mat{1}{-{1 \over r} \um x^+}{{1 \over s}\vm x^-}
   {0}{\um}{0}{0}{0}{\vm}  \eqno(42)$$

\noi where we have introduced the new elements $\um$ and $\vm$ satisfying
$\um u=u\um=\vm v=v\vm=I$. In the Table we have collected all the
information on the quantum group ${\cal M}$ generated by
$I,x^+,x^-,u,v,\um,\vm$.
This group reduces to the matrix group (38) in the classical limit.
The product of two quantum matrices in (41) again satisfies the
exchange relations (35), if we recall that all the elements of $M(x,u,v)$
commute with all the elements of $M(y,w,z)$. On the other hand, the inverse
satisfies the exchange relations (35) with $q \rightarrow \qm$.
\sk

{\sl Note 5}: the non-commutativity of the quantum plane arises as
part of the geometry of the whole Poincar\'e quantum group, rather than
of a representation space for the action of the D=2 rotation group
as in [4,5]. In fact, here the quantum plane has the coset
interpretation \qP / q-Lorentz , the q-Lorentz group in D=2 being the
SO(1,1) generated by $T_0$.
\sk

The functionals $\f{i}{j}$ are also easy to find: they are determined
by their value on the $\M{k}{l}$ (then their value on any power of
$\M{k}{l}$ can be found via the rule (16a)). In
the adjoint representation they
take the form:

$$(f_+)^i_{~j} \equiv \f{+}{+} (\M{j}{i})=
                \Mat{1}{0}{0}{0}{1}{0}{0}{0}{q} \eqno(43a)$$
$$(f_-)^i_{~j} \equiv \f{-}{-} (\M{j}{i})=
                \Mat{1}{0}{0}{0}{\qm}{0}{0}{0}{1} \eqno(43b)$$
$$(f_0)^i_{~j} \equiv \f{0}{0} (\M{j}{i})=
                \Mat{1}{0}{0}{0}{1}{0}{0}{0}{1} \eqno(43c)$$

\noi Note the relation $f_+ f_- = q^{T_0}$.
\sk
The Cartan-Maurer equations of our deformed D=2 Poincar\'e algebra
are :

$$d\om^+ - r \unmezzo \om^+ \we \om^0 = 0$$
$$d\om^- + s \unmezzo \om^- \we \om^0 = 0$$
$$d\om^0=0  \eqno(44)$$

\noi They can be deduced from the general formula [3c]:

$$d \om^k + (T_i T_j)(x^k) \om^i \we \om^j = 0 \eqno(45)$$

Setting

$$\c{k}{ij} \equiv (T_i T_j) (x^k)               \eqno(46)$$

\noi and recalling that $\C{k}{ij}$ can be expressed as

$$\C{k}{ij} \equiv (T_i T_j - \R{kl}{ij} T_k T_l)(x^k)  \eqno(47)$$

\noi (apply (1) to $x^k$ and use (40)) we find the relation:

$$\C{k}{ij} = \c{k}{ij} - \R{lm}{ij} \c{k}{lm} \eqno(48)$$

\noi Thus the structure constants $C$ appearing in the Cartan-Maurer
equations (45) are in general different from the structure constants
${\bf C}$ of the q-commutator algebra (1). In fact, they
coincide only in the
case $\Rh^2 =1$ ($\R{ij}{kl} \R{kl}{rs} = \del^i_r \del^j_s$). For
our q-algebra (30) we have indeed $\Rh^2 =1$, so that its Cartan-Maurer
equations are as given in (44).
\sk
The commutation relations of the ``coordinates" $x^i$ with the left-
invariant one-forms $\om^k$ are deduced from eq. (15), and  wedge
products of $\om^i$ are defined in (23). We list them in the Table
that summarizes the bicovariant calculus on D=2 \qP.
\sk
We point out that other deformations of the $D=2$ Poincar\'e algebra
have been found [13,14]. However, only the one given in (28) leads to a
bicovariant differential calculus.
\sk
We conclude by outlining a procedure to ``gauge" the quantum (Poincar\'e
) groups that follows from the geometric approach advocated in [15] (see
[15b] for a brief introduction).
\sk
The idea is to allow the right-hand side of eq. (45) to be nonvanishing,
and then to interpret it as the curvature $R^k$ associated to
$\om^k$. This is
called ``softening" of the group $G$, and formally the same thing can be
carried out for quantum groups. The closure of the $d$ operator, and the
quantum Jacobi identities (4) lead to differential conditions on the
curvatures, the quantum Bianchi identities:

$$dR^k+\c{k}{ij} R^i \we \om^j - \c{k}{ij} \om^i \we R^j =0 \eqno(49)$$

A quantum Lie derivative [3,12] can be introduced on
the ``soft quantum group",
using its representation as the operator:

$$l_{y}=i_y d + d i_y \eqno(50) $$

\noi that holds also in the quantum case, provided
the contraction operator
$i_y$ is defined appropriately [11] ($y$ belongs to the dual of the soft
$A$, and is the q-analogue of a tangent vector).

This allows the definition of ``quantum diffeomorphisms" :

$$\delta_y \om^k = l_y \om^k = (i_y d + d i_y) \om^k=
        (\nabla y)^k + i_y R^k \eqno(51)$$

\noi where $\nabla$ is the quantum covariant derivative whose definition can
be read off the Bianchi identities (49) $\nabla R^k = 0$.
\sk
The construction of an action invariant under the transformations (51)
proceeds as in the classical case. For example, the lagrangian for
D=4 q-gravity is formally unchanged:

$${\cal L}_q = R^{ab} \we V^c \we V^d \epsi_{abcd}  \eqno(52)$$

\noi $R^{ab}$ refers now to the q-Lorentz part of the q-Poincar\'e
curvature, $V^c$ is the softening of the left-invariant one-form
corresponding to the translations, the wedge product is defined as in
(23), and the $\epsi_{abcd}$ tensor is the quantum generalization of the
alternating tensor for the q-Lorentz group. This of course presupposes
that one can find a q-Poincar\'e group containing q-Lorentz as a
subgroup, which is indeed the case [16, 9].

\sk\sk\sk

\centerline{\bf Table}
\centerline{The bicovariant calculus on D=2 \qP}
\sk\sk
{\sl D=2 \qP Hopf algebra}

$$T_0 T_+ - T_+ T_0 = r T_+$$

$$T_0 T_- - T_- T_0 = -s T_-$$

$$q^{\unmezzo} T_+ T_- - q^{-\unmezzo} T_- T_+ = 0 $$

$$\Dp (T_0)=T_0 \otimes 1 + 1 \otimes T_0$$

$$\Dp (T_\pm)=T_\pm \otimes f_\pm + 1 \otimes T_\pm$$

$$\ep (T_i) = 0; ~~\kp (T_0)=-T_0,~\kp(T_\pm)=-T_\pm~\kp (f_\pm ) $$
\sk

{\sl Adjoint representation}

$$ (T_0)^i_{~j} = -\C{i}{0j} = \Mat{0}{0}{0}{0}{-1}{0}{0}{0}{1}~~~
   (T_+)^i_{~j} = -\C{i}{+j} = \Mat{0}{1}{0}{0}{0}{0}{0}{0}{0}~~~$$
 $$  (T_-)^i_{~j} = -\C{i}{-j} = \Mat{0}{0}{-1}{0}{0}{0}{0}{0}{0}~~~$$

$$(f_+)^i_{~j} \equiv \f{+}{+} (\M{j}{i})=
                \Mat{1}{0}{0}{0}{1}{0}{0}{0}{q}~~~
(f_-)^i_{~j} \equiv \f{-}{-} (\M{j}{i})=
                \Mat{1}{0}{0}{0}{\qm}{0}{0}{0}{1}$$
$$(f_0)^i_{~j} \equiv \f{0}{0} (\M{j}{i})=
                \Mat{1}{0}{0}{0}{1}{0}{0}{0}{1}$$

$$\M{i}{j} = \Mat{I}{{1 \over r}x^+}{-{1 \over s}x^-}{0}{u}{0}{0}{0}{v} $$

$$ \kappa (\M{i}{j}) =
\Mat{1}{-{1 \over r} \um x^+}{{1 \over s}\vm x^-}{0}{\um}{0}{0}{0}{\vm}  $$
\sk

{\sl Exchange relations of the algebra ${\cal M}$}, generated by $I,\xp,\xm,
u,v,\um,\vm$

$$x^+ x^- = q x^- x^+  $$
$$u x^-= q x^- u $$
$$v x^+ = \qm x^+ v $$
$$\vm x^+=q x^+ \vm$$
$$\um x^-=\qm x^- \um$$
\sk
{\sl Hopf structure on ${\cal M}$}
\sk
{}~~~Coproduct:

$$\D (I)=I\otimes I,~~~\D (\xp)=I\otimes \xp + \xp \otimes u,
{}~~~\D (\xm)=I\otimes \xm + \xm \otimes v$$
$$\D (u)=u\otimes u,~~~\D (v)=v\otimes v$$
$$\D (\um)=\um\otimes \um,~~~\D (\vm)=\vm\otimes \vm$$
\sk
{}~~~Counit:

$$\epsi (I)=1,~~\epsi (\xp)=\epsi (\xm)=0$$
$$\epsi (u) = \epsi (v)=\epsi (\um) = \epsi (\vm)=1$$
\sk
{}~~~Coinverse:

$$\kappa (I)=1,~~~\kappa (\xp)=-\um\xp,~~~\kappa (\xm)=-\vm \xm$$
$$\kappa (u)=\um,~~~\kappa (v)=\vm,~~~\kappa (\um)=u,~~~\kappa (\vm)=v$$
\sk
{\sl Bicovariant differential calculus on ${\cal M}$}
\sk
{}~~~Exterior derivative:
$$d\xp=-\xp \om^0 + \om^+,~~~d\xm=-\xm\om^0+\om^-$$
$$du=-ru\om^0,~~~dv=sv\om^0,~~~d\um=r\um\om^0,~~~d\vm=-s\vm\om^0$$
\sk
{}~~~The action of $T_i$ and $\f{i}{j}$:
$$T_\pm (x^\pm)=1,~~T_0 (u)=-r,~~T_0 (v)=s,~~T_0 (\um)=r,~~T_0 (\vm)=-s$$
$$\fp (v)=q,~~\fm (u)=\qm,~~\fp (u)=\fm (v)=\fn (I) =\fn (u)=\fn (v) =1$$
$$\fp (\vm)=\qm,~~\fm (\um)=q,~~\fp (\um)=\fm (\vm)=\fn (\um)=\fn (\vm) =1$$
\sk
{\sl Left and right actions}

$$\DR (\om^i ) = \om^j \otimes \M{j}{i},~~~~\om^i =
{\rm left~invariant~one-forms} $$

$$\DL (\eta^i ) = \kappa (\M{j}{i}) \otimes \eta^j,~~~~\eta^i=
{\rm right~invariant~one-forms} $$

\sk
{\sl Commutation relations of $\M{i}{j}$ and left-invariant forms}

$$\om^+ x^-=q x^- \om^+,~~\om^- x^+= \qm x^+ \om^-$$
$$\om^+ v=q v \om^+,~~\om^- u=\qm u \om^-$$
$${\rm all~ other~ commutations~ are~ trivial}$$
\sk
{\sl Wedge products}
$$\om^+ \we \om^- = \om^+ \otimes \om^- - q \om^- \otimes \om^+$$
$$\om^- \we \om^+ = \om^- \otimes \om^+ - \qm \om^+ \otimes \om^-$$
$$\Rightarrow \om^+ \we \om^-=-q \om^- \we \om^+$$
$${\rm all~ other~ wedge~ products~ as~in~classical~case}$$

\sk
{\sl Cartan-Maurer equations}

$$d\om^+ - r\unmezzo \om^+ \we \om^0 = 0$$
$$d\om^- + s\unmezzo \om^- \we \om^0 = 0$$
$$d\om^0=0  $$

\vfill
\eject

{\bf References}
\sk
\item{[1]} V. Drinfeld, Sov. Math. Dokl. {\bf 32} (1985) 254.
\item{} M. Jimbo, Lett. Math. Phys. {\bf 10} (1985) 63; {\bf 11}
        (1986) 247.
\item{} L.D. Faddeev, N.Yu. Reshetikhin and L.A. Takhtajan, Algebra and
Analysis, {\bf 1} (1987) 178.
\sk

\item{[2]} A. Connes, Publ. Math. IHES Vol. {\bf 62}, 41 (1986);
          ``G\'eometrie non commutative", Inter Editions, Paris (1990).
\sk

\item{[3a]} S.L. Woronowicz, Publ. RIMS, Kyoto Univ., Vol. {\bf 23}, 1
(1987) 117.
\item{[3b]} S.L. Woronowicz, Commun. Math. Phys. {\bf 111} (1987) 613;
\item{[3c]} S.L. Woronowicz, Commun. Math. Phys. {\bf 122}, (1989) 125.
\sk

\item{[4]}Yu.I. Manin, ``Quantum Groups and Non-Commutative Geometry",
 Universit\'e de Montr\'eal, Centre de Recherches Math\'ematiques, 1988.
\sk

\item{[5]} B. Zumino, in ``Recent Advances in Field Theories", Annecy
meeting in honour of R. Stora, 1990;
\item{} J. Wess and B. Zumino, CERN-TH-5697/90.
\item{} W.M. Schmidke, S.P. Vokos and B. Zumino, Z. Phys. C - Particle
and Fields {\bf 48} (1990) 249.
\item{} A. Schirrmacher, J. Wess and B. Zumino, KA-THEP-1990-19
\sk

\item{[6]} U. Carow-Watamura, M. Schlieker and S. Watamura, KA-THEP-1990
-15, to be publ. in Z. Phys. C.
\sk

\item{[7]} U. Carow-Watamura, M. Schlieker, S. Watamura and W. Weich,
     KA-THEP-1990-26.
\sk

\item{[8]} P. Podle\'s and S.L. Woronowicz, Commun. Math. Phys. {\bf 130
} (1990) 381.
\sk

\item{[9]} L. Castellani, ``Noncommutative geometry on the
 quantum Poincar\'e group", Torino preprint, in preparation.
\sk

\item{[10]} T. Curtright and C. Zachos, Phys. Lett. {\bf 243B} (1990)
237.
\sk

\item{[11]} P. Aschieri and L. Castellani, ``An introduction to
non-commutative differential geometry on quantum groups", Torino
preprint, in preparation.

\sk

\item{[12]} D. Bernard, in the notes of the E.T.H workshop, Z\"urich
1989.
\sk

\item{[13]} E. Celeghini, R. Giachetti, E. Sorace and M. Tarlini,
``Contractions of quantum groups", Proceedings of the semester ``Quantum
Groups", Euler Math. Institute Leningrad, Oct.-Nov. 1990; J. Math. Phys.
{\bf 32} (1991) 2548; {\sl ibidem} {\bf 31} (1991) 2548.
\sk
\item{[14]} S.L. Woronowicz, ``Quantum SU(2) and E(2) groups.
   Contraction procedure.", Claude Bernard University preprint,
Lyon (1991).
\sk
\item{[15a]} L. Castellani, R. D'Auria and P. Fr\'e,
{\sl ``Supergravity and
Superstrings: a geometric perspective"}, World Scientific (1991),
Singapore.
\item{[15b]} L. Castellani, ``Group
geometric methods in supergravity and
superstring theories", Lectures presented at the C.B.P.F, Rio de Janeiro
1990 (to be publ. in Int. Jou. Mod. Phys. A).
\sk
\item{[16]}V. Dobrev, private communication.

\bye